## Graphic Abstract

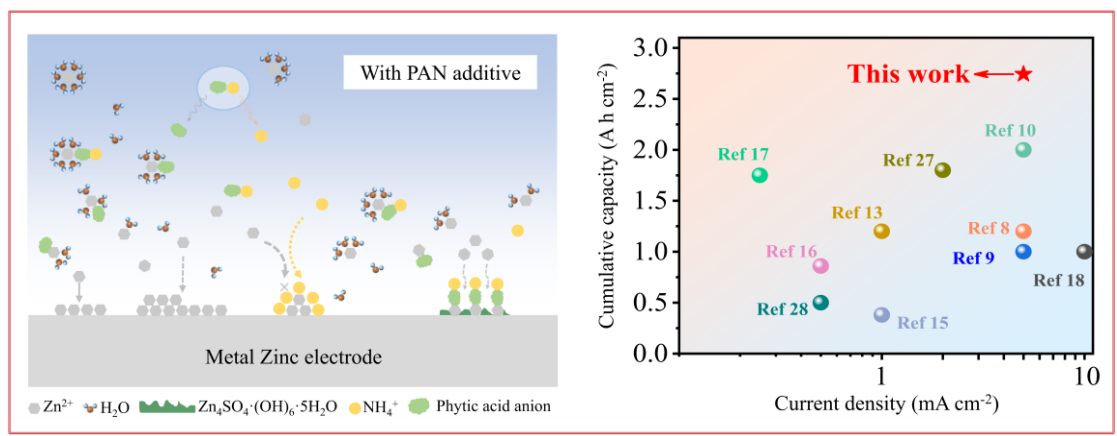

**Caption of Graphic Abstract**

By introducing a minute quantity of phytate ammonium into the conventional dilute zinc sulfate electrolyte, a static physical barrier is established by the phytic anion. Simultaneously, a dynamic electrostatic shield layer collaborates, yielding superior electrochemical performance. This addition facilitates an accelerated de-solvation process, effectively suppressing the so-called "tip effect" and mitigating dendrite growth.

# Zincophilic armor: Phytate ammonium as a multifunctional additive for enhanced performance in aqueous zinc-ion batteries

Fangyuan Xiao[a,b], Xiaoke Wang[a,b], Kaitong Sun[a], Qian Zhao[a], Cuiping Han[b,*], and Hai-Feng Li[a,*]

[a]*Institute of Applied Physics and Materials Engineering, University of Macau, Avenida da Universidade, Taipa, Macao SAR 999078, China*
[b]*Faculty of Materials Science and Energy Engineering/Institute of Technology for Carbon Neutrality, Shenzhen Institute of Advanced Technology, Chinese Academy of Sciences, Shenzhen Guangdong 518055, China.*

*Email addresses:* cp.han@siat.ac.cn (C. Han)
                              haifengli@um.edu.mo (H.-F. Li)

Corresponding author: haifengli@um.edu.mo (H.-F. Li),
Full postal address:
IAPME, N23-3002, University of Macau
Avenida da Universidade, Taipa, Macao SAR 999078, China
Phone: +853 6307 5422

## Abstract
Corrosion and the formation of by-products resulting from parasitic side reactions, as well as random dendrite growth, pose significant challenges for aqueous zinc-ion batteries (AZIBs). In this study, phytate ammonium is introduced into the traditional dilute Zinc sulfate electrolyte as a multi-functional additive. Leveraging the inherent zincophilic nature of the phytic anion, a protective layer is formed on the surface of the zinc anode. This layer can effectively manipulate the deposition process, mitigate parasitic reactions, and reduce the accumulation of detrimental by-products. Additionally, the competitive deposition between dissociated ammonium ions and $Zn^{2+}$ promotes uniform deposition, thereby alleviating dendrite growth. Consequently, the modified electrolyte with a lower volume addition exhibits superior performance. The zinc symmetric battery demonstrates much more reversible plating/stripping, sustaining over 2000 hours at 5 mA cm$^{-2}$ and 1 mA h cm$^{-2}$. A high average deposition/stripping efficiency of 99.83% is achieved, indicating the significant boosting effect and practical potential of our strategy for high-performance aqueous zinc-ion batteries.

# 1. Introduction

Lithium-ion batteries (LIBs) have found widespread applications since their initial commercialization by SONY in 1991 [1]. The expanding market scale has led to rapid resource consumption, resulting in rising costs. Additionally, traditional commercial batteries often utilize organic compounds as solvents paired with polymer separators, posing a hidden danger of fire in the event of a short circuit or overheating. This risk it particularly pronounced when batteries are used in electric vehicles or other high-power facilities through series connection [2, 3]. Under extreme conditions, local overheating caused by a short circuit can lead to the decomposition of the organic electrolyte. As the temperature continues to rise, the battery may undergo firing and combustion, posing a threat to personal safety [4]. Zinc-ion batteries (ZIBs) emerge as promising candidates for the next generation of large-scale energy storage devices, offering a potential solution to the challenges faced by LIBs. Zinc metal boasts a high specific capacity (820 mA h g$^{-1}$) and an appropriate reduction potential (Zn$^{2+}$/ Zn, -0.76 V vs. SHE), providing favorable compatibility with aqueous electrolyte [5-7]. By utilizing water as the electrolyte solvent, battery security can be greatly improved compared to traditional organic solvent electrolytes. However, challenges persist with aqueous electrolytes.

Traditional aqueous zinc sulfate (ZnSO$_4$) electrolytes suffer from several typical problems, which can be categorized into three parts: i) Dendrite growth caused by random zinc ion flux and deposition [8-11], significantly reducing the cell's lifespan. ii) By-product accumulation resulting from unexpected side reactions, where by-products typically consist of zinc hydroxyl sulfate (ZHS, Zn$_4$SO$_4$(OH)$_6$·nH$_2$O). The accumulation of ZHS increases the interface resistance, hindering zinc ion transport and decreasing electrode reaction kinetics [12-14]. iii) Weakly acidic electrolytes inevitably corrode the metal anode and aggravate hydrogen evolution reactions (HER) simultaneously [15-18]. These facts result in extra metal consumption due to irreversible reactions, leading to the rapid degradation of the battery. Furthermore, continued corrosion and metal consumption generate gas without pressure relief and buffer, posing a danger of excessive flammable and explosive hydrogen gas accumulation in the sealing system. Additionally, some zinc ions tend to combine with H$_2$O to form hydrated zinc ions (Zn(6H$_2$O)$^{2+}$) through a solvation process in a solution system. The larger ion size of Zn(6H$_2$O)$^{2+}$ compared to bare Zn$^{2+}$ shows down ion transference [19-22], and water molecules impede the de-solvation process, resulting in slower ion transport. The competitive reduction between Zn$^{2+}$ and H$^+$ leads to severe side reactions with an acid electrolyte [23, 24]. These issues must be addressed before the practical application of AZIBs.

Numerous strategies have been proposed to address these challenges, including electrode modulation [25], electrode-electrolyte interface engineering [26], separator design [27], and electrolyte optimization [28, 29]. These approaches aim to regulate ion solvation structure and random deposition, thereby mitigating dendrite growth and unfavorable HER. Functional additives are garnering increasing research interest due to their low cost, ease of implementation, and high effectiveness [30-32]. Among various functional additives, organic molecules represent an important category [33-35]. For instance, Xie et al. [36] proposed polyacrylic acid (PAA) as an additive for a sol-containing electrolyte, resulting in the formation of an in-situ solid electrolyte interface and achieving exceptional battery life. Additionally, Sun [37] introduced MXene nanosheets into a typical electrolyte, leading to an enhanced lifespan through facilitated ion migration.

Ammonium ions have been proven to efficiently regulate the deposition behavior of $Zn^{2+}$ by forming a dynamic electrostatic shielding layer on the zinc metal surface [38]. Ammonium ions absorb at the tip/raised part of the metal surface, preventing subsequent vertical deposition of $Zn^{2+}$ through their static shielding effect on the tip area. This slows down irregular deposition speed, ensuring a uniform deposition process. Phytic acid serves as a typical metal corrosion inhibitor and organic ligand, significantly improving the electrochemical performance of the zinc anode after treatment [39]. The improved performance results from the establishment of an artificial protection layer containing phytate zinc. Additionally, a special ion transfer mechanism is prompted, where a special electro-migration channel formed by phytic anion and zinc ion acts as an efficient pathway for fast ion transmission. The deposition energy barrier of zinc ions is reduced with surface modulation.

In this study, we developed an electrolyte engineering strategy using a multifunctional additive of phytate ammonium (PAN). On one hand, owing to the inherent zincophilicity of the phytic anion, a physical barrier composed of zinc salt with phytic is generated. This layer enhances the stability of the electrode-electrolyte interface, mitigates side reactions, and reduces the accumulation of by-products. On the other hand, the static electric shield of $NH_4^+$ suppresses the "tip effect" (dendrite growth from uneven deposition of zinc ions caused by the inhomogeneous distribution of the electric filed) during the deposition process, achieving uniform deposition. With a small amount of PAN addition, the cyclic performance of diluted $ZnSO_4$ is remarkably improved. Benefiting from the mentioned merits, the lifespan of the Zn‖ Zn symmetric battery improves nearly tenfold, from 155 hours to 1500 hours, with 1% volume multifunctional additive PAN at 1 mA cm$^{-2}$, 1mA h cm$^{-2}$. A high average Coulombic efficiency of 99.83% is achieved within 2000 cycles of Zn‖ Cu asymmetry battery, indicating favorable reversibility and cycle stability. A full battery paired with a typical $V_2O_5$ cathode also exhibits acceptable capability, illustrating the practical application potential of this additive design.

## 2. Experimental

### 2.1. Materials

Phytic acid ($C_6H_{18}O_{24}P_6$, solution in water, 50 wt%), N-Methylpyrrolidone (NMP, 99.9%), and Zinc sulfate heptahydrate ($ZnSO_4 \cdot 7H_2O$, AR) were purchased from Shanghai Aladdin Bio-chem Technology. Ammonia ($NH_3 \cdot H_2O$, 25% ~28%) was purchased from SINOPHARM, and Vanadium pentoxide ($V_2O_5$, 99.6%) was acquired from Alfa Aesar. Ketjen black was obtained from SUZHOU SINERO TECHNOLOGY, and PVDF 5120 was purchased from Guangdong Canrd New Energy Technology.

### 2.2. Electrolyte/ Electrode preparations

1M $ZnSO_4 \cdot 7H_2O$ solution was prepared in deionized water to create the pristine electrolyte, labeled as the 'bare electrolyte' (denoted as BE). For the additive, 65 μL of Phytic acid and 50 μL of Ammonia were added to 5 mL of deionized water, resulting in an additive denoted as 'PAN', with a pH value around 5.5. The modified electrolyte containing 1% PAN was prepared by adding a 1% volume ratio PAN solution to the bare electrolyte. Additionally, a series of electrolytes with varying PAN concentrations (0.5%, 2%, and 5% by volume) were prepared using the same procedure. All preparation steps were conducted at room temperature.

Commercial zinc foil with different thicknesses (10 μm, 20 μm, and 80 μm) was purchased, polished with sandpaper, and wiped with ethanol to remove the oxide layer and impurities. The processed zinc foil was cut into small pieces measuring 1 cm × 1 cm, which were used as electrodes for the symmetric battery and anode for the full battery. To prepare the cathode, commercial $V_2O_5$ (without any modification), Ketjen black, and PVDF were mixed with a mass ratio of 7: 2: 1 using an agate mortar, with NMP as the dispersant. The uniform slurry was then coated on a stainless steel (SS, 100 μm) foil with a surgery blade. After evaporating the dispersant in an oven at 80 ºC for 12 hours, the resulting cathode was obtained, featuring an average mass loading of ~ 1 mg cm$^{-2}$ and an N/P ratio of ~ 28.4.

### 2.3. Characterizations

Microstructure images were obtained using a Scanning Electron Microscope (Zeiss Sigma FESEM) equipped with Energy Dispersive Spectroscopy (EDS, Oxford) for element analysis. Crystalline data of the electrodes were measured by X-ray diffractometer (Rigaku, SmartLab 9 kW), with copper as the anode target material and output Kα radiation at 1.5418 Å (containing Kα$_1$ = 1.54056 Å and Kα$_2$ =1.54439 Å as the radiation with an intensity ratio of 2: 1). The two-theta degree range was from 5 to 60$^o$ with a scan rate of 10 $^o$/min. The contact angle of the electrolyte with the metal foil was tested by a Drop Shape Analyzer (Ningbo Scientific Instruments, OSA100S-T-P). Binding energy information of the functional group with the electrode was detected by X-ray photoelectron spectroscopy (Thermo Fisher Scientific, ESCALAB250Xi).

*2.4. Electrochemical measurements*

Coin cells were assembled in CR2032 type, with zinc foil and glass fiber (GF/D, Whatman) used as anode and separator, respectively. Commercial $V_2O_5$ served as the cathode material for the full battery, with a 1M $ZnSO_4$/1M $ZnSO_4$+1% PAN electrolyte.

Long cycle performance was measured using the Neware battery test system (Shenzhen, China). Coulombic efficiency (CE) was tested by a Zn‖ Cu asymmetric battery, used to estimate the plating/ stripping efficiency. Cyclic Voltammetry (CV), Chronoamperometry (CA), Liner Polarization (LP) test, and Electrochemical Impedance Spectroscopy (EIS) tests were analyzed using an Electrochemical Workstation (CH Instruments, CHI760E, Shanghai). For the CV test, the scan rate was 1 mV s$^{-1}$ with a voltage range from 0.2 to 1.9 V. CA curves were recorded with a -150 mV constant potential for 400 seconds. The potential range of Liner polarization was -0.2 V to 0.2 V with a scan rate of 1 mV s$^{-1}$. EIS was probed by the electrochemical workstation at a frequency range from 10 mHz to 100 kHz with an amplitude of 5 mV. All these tests were performed with coin cells at room temperature unless specially marked.

# 3. Results and discussion

*3.1. Influence of PAN on the morphology of zinc anode*

Fig. 1**a** presents a schematic illustrating some typical challenges associated with traditional dilute $ZnSO_4$ electrolytes. While the zinc metal anode boasts a high theoretical capacity, several difficulties hider the widespread application of zinc-ion batteries on a large scale. These challenges include the slow kinetics of the de-solvation process from zinc hydrates to zinc ions, the formation of dendrites during random and uneven zinc ion deposition, surface passivation resulting from the accumulation of inert by-products, and simultaneous corrosion and hydrogen evolution reactions. Upon the addition of PAN, the cations and anions decomposed from PAN enhance the electrochemical performance in several ways: the phytic anion actively participates in the solvation process of zinc ions, adjusting the solvation structure for fast ion transportation. Simultaneously, a protective layer is formed, leveraging the intrinsic zincophilic properties of the phytic anions on zinc metal. This in-situ protection layer,

composed of phytate zinc, regulates ion flux for a uniform deposition process. Additionally, the additive mitigates side reactions and reduces by-product accumulation, alleviating random zinc deposition and dendrite growth by restraining the "tip effect" from ammonium ions. In Fig.1**b**, the contact angle test results show that the pristine electrolyte without the addition has a contact angle of 85.6º, which decreases to 80.5º after the addition of PAN. This indicates that the additive improves wettability to the electrode, facilitating faster infiltration and contributing to the construction of a stable solid electrode interface, resulting in refined zinc crystal size during the deposition process. The surface morphology evolution of the electrode during aging is illustrated in Fig. 1**c**. After 20 hours (equivalent to 10 cycles at 1 mA cm$^{-2}$, 1 mA h cm$^{-2}$), the electrode demonstrates a relatively flat and uniform surface. In contrast to the smooth surface of the zinc metal foil (Fig. S1**a**), micro-clusters and microspheres are observed on the zinc electrode when paired with the pristine electrolyte. These features are believed to be by-products of chemical corrosion (Fig. S1**b**). The superior anode consistency in the aging test demonstrates that the additive imparts better anti-corrosion ability to the zinc anode. At high magnification in the SEM images of the zinc electrode with the additive (Fig. S1**c**) and corresponding EDS mapping results (Fig. S1**d**), a dual defense mechanism is hypothesized to be established. This mechanism is derived from the synergistic function of phytic anions and ammonium ions: a protection layer formed by phytic anions and a dynamic electrostatic shield layer from the superior absorption of $NH_4^+$. The uniform distribution of C and N elements on the metal surface suggests the presence of an in-situ protection layer. Cross-section images in Fig. S1**e** and Fig. S1**f** confirm the existence of this thin protection layer and its positive effect on electrode protection. Dissociated ammonium ions suppress the tip effect, extending the lifespan. X-ray diffraction results with the electrode from the aging battery support this conclusion. Besides the strong characteristic peaks from zinc metal after 35º, a single peak appears at 8.09º with both electrodes after aging (Fig. 1**d**), attributed to ZHS ($Zn_4SO_4(OH)_6\cdot5H_2O$, JCPDS No. 78-0246). Due to the protection layer on the zinc metal anode surface, the intensity of this characteristic peak decreases after the addition of PAN, suggesting that PAN effectively reduces side reactions and by-product accumulation. To further confirm the influence of the additive, X-ray photoelectron spectroscopy (XPS) of the electrodes was surveyed, and the results are exhibited in Fig. 1**e**. The signals observed at 286.19 eV for C 1s and 532.37 eV for O 1s are attributed to the presence of the C-O bond originating from the phytic anion. The existence of the C-O bond on the electrode surface reveals that the additive is absorbed on the zinc metal surface due to its zincophilicity, consistent with the EDS mapping results in Fig. S1**d**. Additionally, a pair of S 2p peaks is detected in both samples (Fig. S2), illustrating that the cell still experiences some degree of side reactions, mainly caused by chemical corrosion.

*3.2. Electrochemical performance improvement*

As shown in Fig. 2**a**, we assembled a Zn∥ Zn symmetric battery for chronoamperometry test. The bare ZnSO$_4$ electrolyte exhibits a prolonged time to achieve a stable current density, typically described as a two-dimensional (2D) diffusion behavior, as illustrated in the insert scheme in Fig. 2**a**. In this scenario, zinc ion flux and deposition concentrate on a 2D plane, leading to vertical growth and the formation of dendrites that can damage the separator, potentially causing short circuits. After adding the PAN additive, the corresponding current density stabilizes in less than 50 seconds, attributed to a three-dimensional diffusion model. Zinc ions preferentially deposit on the surface in a parallel mode, promoting uniform deposition and reducing dendrite formation. The liner polarization test results depicted in Fig. 2**b** illustrate the anodic/cathodic polarization process, simulating the Tafel slope and calculating the corrosion current for different electrolytes. It is observed that the corrosion current decreases from 0.87 mA to 0.51 mA after adding of PAN. Although the additive is weakly acidic, the corrosion potential remains basically unchanged, confirming that the additive does not exacerbate corrosion and hydrogen evolution reaction (HER). Cyclic voltammetry of the symmetric battery in Fig. S3**a** depicts a lower overpotential after addition, revealing reduced resistance and energy barrier, facilitating an easier zinc ion deposition process. The electrochemical impedance spectra of the symmetry battery paired with electrolytes containing varying amounts of the additive are illustrated in Fig. S3**b**, and S3**c**. The lowest charge transfer resistance is observed with a 1% volume addition, establishing it as the optimal amount for subsequent research. At 1 mA cm$^{-2}$ and an area capacity of 1 mA h cm$^{-2}$, the symmetry battery with the bare electrolyte fails within 155 hours (Fig. 2**c**). In contrast, the electrolyte with the additive demonstrates stable cycling performance for more than 1500 hours, with a voltage hysteresis of only 13.9 mV after 1200 hours, indicating high reversibility and an efficient stripping/deposition process. The rating performance of the symmetric battery in Fig. 2**d**, applied at 1 mA h cm$^{-2}$ with series current densities of 1, 2, 5, 10, and 20 mA cm$^{-2}$, respectively, maintains stable cycling performance even at high current densities, with the voltage hysteresis at different current densities lower than that without the additive, verifying the improvement of PAN addition to rating performance. A Zn∥ Cu asymmetric battery is typically assembled to assess Coulombic efficiency (CE). In Fig. 2**e**, Coulombic efficiency of bare ZnSO$_4$ electrolyte fluctuates 100 % after 500 cycles, with a distinct fluctuation observed around 1000 cycles (see insert picture on Fig. 2**e**). In contrast, the electrolyte with the additive maintains good reversibility even after 2000 cycles, confirming that PAN addition enables a much more stable deposition process.

The results reveal a high initial Coulombic efficiency (97.3%) and a higher average Coulombic efficiency of 99.83% after 2000 cycles, superior to many former reports under the same condition (Fig. 2**f**) [40, 41]. Compared with the bare ZnSO$_4$ electrolyte (Fig. 2**g**), the nucleation overpotential (NOP) decreases to 16.6 mV after the addition of PAN. The lowered NOP indicates a lower nucleation barrier and an improved interfacial activity, suggesting that PAN addition allows for better nucleation kinetics. Additionally, as shown in the voltage capacity curve in Fig. 2**h** and Fig S4, the polarization overpotential profit reduces from 91 mV to 74 mV and maintains a similar value in the first 100 cycles. Well-overlapped voltage curves also indicate good anode stability and reversibility. The electrochemical performance test results above imply the effective function of the PAN additive in implementing a better lifespan of the zinc anode.

Further investigation reveals the enhanced performance of the electrolyte with PAN additive. The Zn|| Zn symmetric battery demonstrated an extremely stable cyclic performance (over 2000 hours) at a current density of 5 mA cm$^{-2}$, 1mA h cm$^{-2}$. It also exhibited a low voltage hysteresis during the long-term charge and discharge process (Fig. 3**a**), attributed to the adjustment of the solvation structure of Zn$^{2+}$ and a more ordered ion flux. In contrast, the symmetric battery with bare ZnSO$_4$ electrolyte suffered a short circuit at 190 hours due to a large over-potential, water involvement side reactions, and the formation of low ionic conductivity by-products, leading to a faulty anode surface and rapid battery failure. Depth of discharge (DOD) is an important standard reflecting the utilization of the metal anode, with a higher DOD approaching practical conditions. When the DOD increased to 21.3%, the cell smoothly ran for over 1000 hours with a lower over-potential, as Figure. S5**a** shows. At a large DOD of 42.6% in Fig. 3**b**, the battery could still maintain steady performance for more than 160 hours. Even at a specific high level to 51.1%, a stable cycle of more than 150 hours could also be realized (Fig. S5**b**), providing the practical value of our design. As Fig. 3**c** displays, the as-assembled coin cell obtained a cumulative capacity of more than 2700 mA h cm$^{-2}$ at the current density of 5 mA cm$^{-2}$ [42, 43]. The improvement in cumulative capacity in this work is remarkable compared with other studies.

### 3.3. Alleviated the accumulation of by-products and enhanced zinc ion dynamics

To explore the instant influence of the PAN addition on the charge/discharge process, morphology, and crystalline details of the electrode after short-term cycles are gathered. From the X-ray diffraction results (Fig. 3**d**) of the electrode after ten cycles, compared with the Zinc plate paired with PAN additive, the electrode with bare electrolyte shows a stronger characteristic peak of ZHS at 8.09 $^{\circ}$ than that with the additive. The physical barrier built with the phytic anion could alleviate the accumulation of $Zn_4SO_4(OH)_6 \cdot 5H_2O$ to a certain extent and regulate an even $Zn^{2+}$ ion flux. Furthermore, after 10 cycles, optical image (inserted in Fig. S6**a**, S6**b**) also reveal a noticeable difference: a darker and more homogeneous electrode surface was obtained after the addition, indicating modification of the electrode surface. Messy deposited Zn and a large amount of ZHS could be noticed on the Zn plate surface (Fig. S6**a**). This is in stark contrast to the neat and flat surface of the electrode paired with the modified electrolyte (Fig. S6**b** and S6**c**), demonstrating the positive effect of the PAN additive in ameliorating the accumulation of ZHS. XPS results in Fig. 3**e** evidenced a similar result: the S 2p orbit splitting signal on 168 eV is detected on the Zn plate paired with BE electrolyte after 10 cycles. However, there is no obvious response of the S 2p orbit splitting on the electrode surface after addition, turned out a low-level accumulation of ZHS than the bare electrolyte. Additionally, a C-O bond could be found on the fine spectrum of C 1s and O 1s after the cycle (Fig. S7**a**), which is in consistent with previous results of the aged battery of a protection layer hypothesis. In addition, nitrogen signals of N 1s are also discovered from the aged and cycled battery with PAN additive (Fig. S7**b**). It is believed that the corresponding peak belongs to $NH_4^+$ of the residue of PAN. The corresponding N 1s peak is detected in the addition sample before the cycle, too. This further confirms the existence of the protection layer. Furthermore, according to previous reports [38], there exists a competitive absorption phenomenon between dissociated $NH_4^+$ and $Zn^{2+}$ during the deposition process. $NH_4^+$ tends to be preferentially absorbed onto the surface, thereby forming an electrostatic shield on protuberant areas. Electrostatic repulsion will stop the subsequent deposition and compel zinc ions to plate in the nearby area, which could effectively inhibit the vertical deposition of zinc ions and then alleviate the random dendrite growth. Ammonium phytate will also undergo partial dissociation in solution. It is speculated that the static electric shield layer containing $NH_4^+$ has formed in the deposition process, too, promoting reversibility and lifespan.

Electrochemical impedance spectroscopy (EIS) was also utilized to elucidate the migration of zinc ions during the charge and discharge process. As shown if Fig. 3**f**, using the inert picture as a reference model, the electrochemical impedance of the symmetric battery with 1M $ZnSO_4$ as the electrolyte was measured to be 425 Ω. Upon the addition of 1% vol PAN, the electrochemical impedance prior to cycling decreased to 100 Ω. Faster ion transfer and better reaction kinetics are obtained with the functional addition. By combining the Chronamperometry (CA) technique with electrochemical impedance spectroscopy testing, we can estimate the ion transference number of the symmetric battery with different electrolytes [44]. The ion transference number follows:

$$T = \frac{I_S(\Delta V - I_0 R_0)}{I_0(\Delta V - I_S R_S)}, \quad (1)$$

where T represent the zinc ion transference number, $I_0$ is the initial current, $I_s$

corresponds the steady-state current, $R_0$ and $R_s$ refer to the resistance of the initial and steady states, $\Delta V$ is the amplitude (applied with 5 mV). The EIS results before and after the addition are depicted in Fig. 3**g**, where the electrolyte with the functional additive delivers a lower charge transfer resistance of 117.4 $\Omega$ compared to that with the bare electrolyte (412.8 $\Omega$) based on the model (insert in Fig. 3**f**) simulation results. Calculated with the equation above, the zinc ion transference number increased from 0.41 to 0.595 after the addition. A higher ion transference number represents a more rapid ion migration, which is indispensable for a uniform electric field distribution and even deposition of zinc ions. The result confirms a more efficient ion transport with the adjustment of the solvation structure from the phytic anion. Activation energy ($E_a$) is an important parameter that reflects the de-solvation energy barrier [45]; it can be calculated with the:

$$\ln \frac{1}{R_{ct}} = \frac{E_a}{RT}, \tag{2}$$

where $R_{ct}$ represents the charge transfer resistance, $E_a$ represents activation energy, R = 8.314 J/(mol•K) is the gas constant, and T is temperature in Kelvin. EIS data from 10 ºC to 60 ºC is presented in Fig. 3**g** and Fig S8, and the cell with a modified electrolyte exhibits lower charge transfer resistance at each temperature condition. There is a liner relationship between ln (1/ $R_{ct}$) and 1/ T (Fig. 3**h**). Calculations with the formula yield the activation energy results. The activation energy of the bare electrolyte is 23.82 kJ mol$^{-1}$, it reduces to 15.35 kJ mol$^{-1}$ after adding the additive, indicating a lower de-solvation barrier and faster de-solvation process, attributed to the weakened solvation effect of $H_2O$ after the phytic anion addition. Phytic ion and ammonium ion work synergistically to achieve a faster, uniform ion flux, well-distributed deposition of Zinc ions, and mitigated side reaction.

*3.4. Practical application*

A full battery with the typical cathode material vanadium oxide ($V_2O_5$) is assembled to further evaluate the effect of the PAN additive in a practical scale. Bare 1M $ZnSO_4$ and that with the additive serve as electrolyte for the full battery, respectively. The cyclic voltammetry (CV) results (first three cycles) of the assembled full battery are displayed in Fig. 4**a** and Fig. S9. Two pairs of redox peaks are observed, with two cathodic peaks at 0.9 V and 0.45~ 0.5 V marked as $C_1$ and $C_2$, respectively, attributed to the co-insertion of $Zn^{2+}$/ $H^+$, and corresponding anodic peaks marked as $A_1$ and $A_2$ represent the extraction of $Zn^{2+}$/ $H^+$. Solid electrolyte interface (SEI) formation in the first cycle results in higher polarization, then the polarization potential of both redox peaks decreases. As the cycles continue, the reaction tends to be more stable and reversible, as illustrated by overlapping curves. The long-term cyclic performance of Zn‖ $V_2O_5$ full battery is displayed in Fig. 4**b**, where the discharge capacity of the commercialized $V_2O_5$ cathode reaches 115.2 mA h g$^{-1}$ in the first cycle. As the cell activates gradually, the specific capacity rises to 167.1 mA h $g^{-1}$ once and remains 65.9 mA h $g^{-1}$ after 200 cycles, achieving a high Coulombic efficiency of 99.8 %. Galvanostatic charge and discharge (GCD) profiles measured at 2 A $g^{-1}$ show a charge plateau around 1.24 V and a discharge plateau at nearly 0.89 V in the early cycles, with another discharge plateau around 0.52 V detected in subsequent cycles, consistent with the results of CV curves. The rating capability for the full battery with the PAN additive is shown in Fig. 4**d**, with an average charge capacity of 102.8, 106.5, 83.3, and 60.6 mA h $g^{-1}$ when applied at 1.0, 2.0, 5.0, and 10 A $g^{-1}$, respectively. As the current density is set back to 2.0 A $g^{-1}$, the specific capacity recovers to 113.3 mA h $g^{-1}$ without obvious capacity fade, demonstrating exceptional performance and reversibility at high current density. Electrochemical impedance spectroscopy of the full battery before cycling are displayed in Fig. 4**e**. According to the model (inserted at Fig. 4**e**, where $R_1$ refers to solution resistance, CPE represents Constant Phase Angle Element, $R_2$ is charge transfer resistance, and Wo is the Warburg Element) simulation results, charge transfer resistance ($R_{ct}$) of the full battery has decreased from 100.4 $\Omega$ to 27.86 $\Omega$ after the addition, illustrating the acceleration of ion transference and better reaction kinetics with the small amount additive.

## 4. Conclusions

In summary, we have proposed a simple and feasible strategy to enhance the stability of the zinc anode when applied with a traditional dilute $ZnSO_4$ electrolyte. By harnessing the intrinsic chelating ability of the phytic anion and its affinity for $Zn^{2+}$, we modified the solvation structure of $Zn^{2+}$ and achieved a uniform deposition process, thereby significantly enhancing ion transport efficiency. Furthermore, the restrain of tip effects from $NH_4^+$ could alleviate dendrite growth. Benefiting from the synergistic effect of the additive, the electrochemical performance of the dilute $ZnSO_4$ electrolyte has been effectively enhanced. The Zn‖ Zn symmetric battery sustained a long cyclic lifespan of more than 2000 hours at 1 mA h $cm^{-2}$ and exhibited a high average plating/stripping Coulombic efficiency of 99.83%. It also demonstrated superior cyclic stability for more than 150 hours at a high DOD of 51.1%. The full battery with $V_2O_5$ displayed acceptable cycle performance and stability. We believe our findings could provide a reference and inspire future research in the field of aqueous zinc-ion batteries.

## CRediT authorship contribution statement
**Fangyuan Xiao:** Conceptualization, Methodology, Investigation, Data curation, Writing – original draft, Validation, Visualization. **Xiaoke Wang:** Investigation, Data curation. **Kaitong Sun:** Investigation, Data curation. **Qian Zhao:** Investigation, Data curation. **Cuiping Han:** Conceptualization, Methodology, Writing – review & editing,

Funding acquisition, Visualization, Supervision. **Hai-Feng Li:** Conceptualization, Methodology, Writing – review & editing, Funding acquisition, Visualization, Supervision.

**Declaration of Competing Interest**

The authors declare that they have no known competing financial interests or personal relationships that could have appeared to influence the work reported in this paper.

**Data availability**

Data will be made available on request.

**Acknowledgements**

The work at Shenzhen Institute of Advanced Technology was supported by Shenzhen Science and Technology Program (Grant No. RCYX20221008092934093). The work at University of Macau was supported by the Science and Technology Development Fund, Macao SAR (File Nos. 0090/2021/A2 and 0049/2021/AGJ), University of Macau (MYRG2020-00278-IAPME), and the Guangdong-Hong Kong-Macao Joint Laboratory for Neutron Scattering Science and Technology (Grant No. 2019B121205003).

**Appendix A. Supplementary data**

Supplementary data associated with this article can be found in the Elsevier Library or from the author. Supplementary data to this article can be found online at

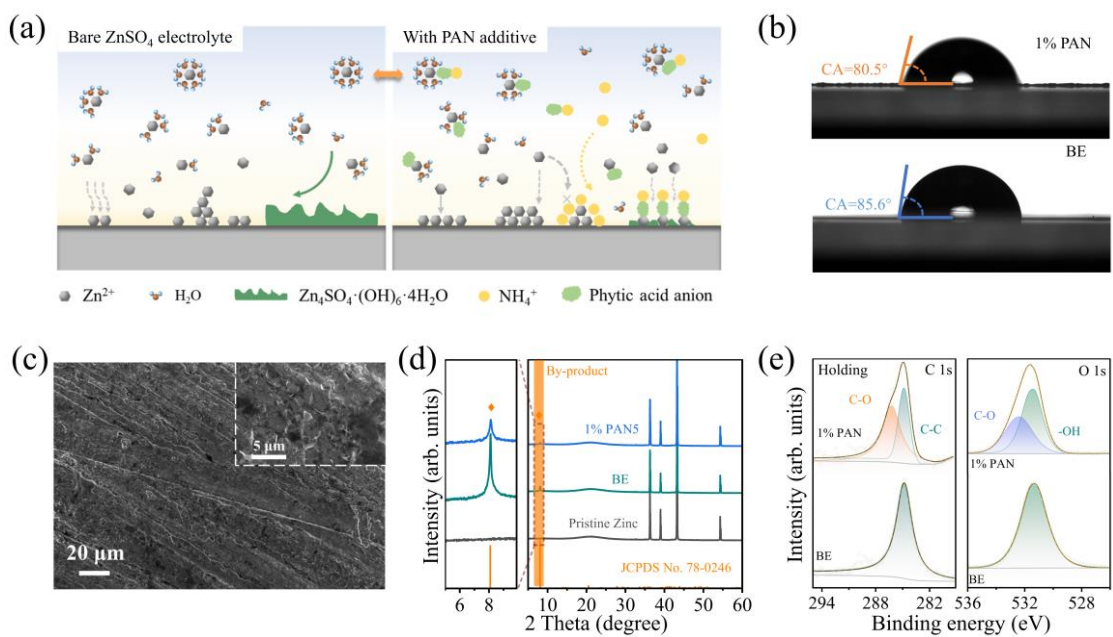

**Fig. 1**. a) Schematic of the primary challenges for the study of zinc-metal anode and function of the PAN additive. b) Contact angle test of bare electrolyte with and without the additive. c) SEM images of the zinc-metal anode with PAN after aging for 20 hours. d) XRD patterns of the zinc-metal anode before and after aging, e) Fine spectra from XPS survey around C 1s and O 1s edges of BE and with the PAN additive after aging.

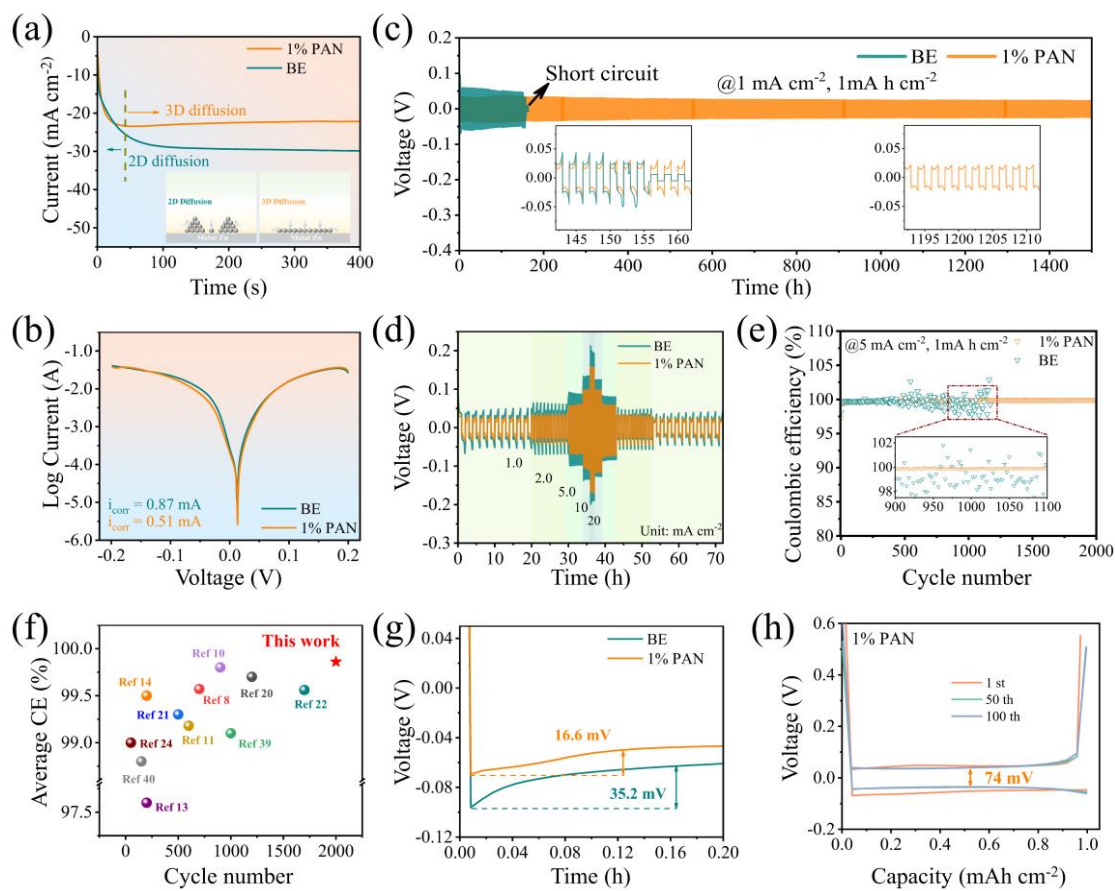

**Fig. 2**. Electrochemical performance. a) Chronoamperometry test for the diffusion behavior of zinc ions. b) Tafel plots of electrolytes and the calculated corrosion current. c) Long-term performance of Zn∥Zn symmetric batteries on 1 mA cm⁻², 1 mAh cm⁻². d) Rating performance with a current-density range from 1 to 20 mA cm⁻² and e) Coulombic efficiency with a Zn∥Cu asymmetric battery on 5 mA cm⁻², 1 mAh cm⁻². f) A comparison of the average Coulombic efficiency between our results and those reported in previous literature. g) Nucleation overpotential of Zn²⁺ at 1 mAh cm⁻². h) Voltage-capacity curves at different cycles with the PAN additive.

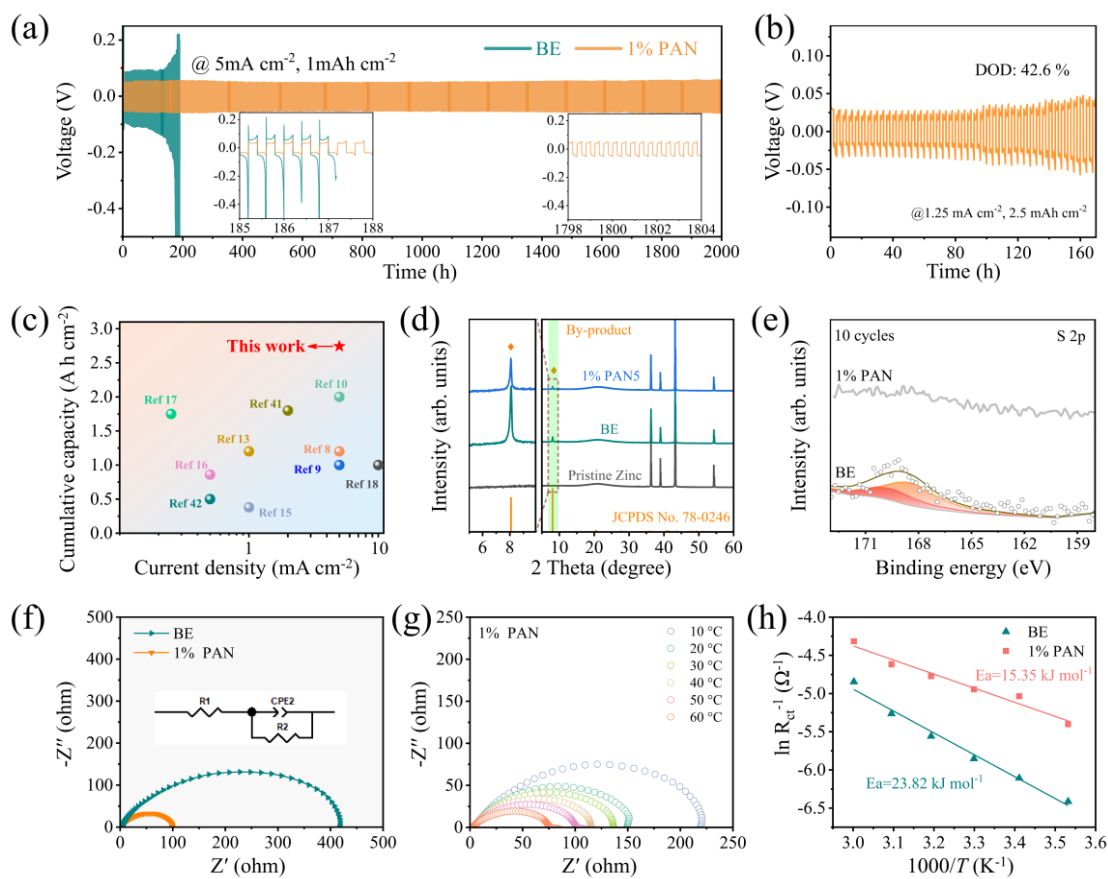

**Fig. 3**. a) Long cyclic test at 5 mA cm⁻², 1 mAh cm⁻². b) Depth of discharge test on 1.25 mA cm⁻², 2.5 mAh cm⁻² with a symmetric battery. c) A comparison of cumulative capacity with results reported in previous studies. d) XRD patterns of the zinc metal anode after 10 cycles at 1 mA cm⁻². e) Fine spectra of S 2p of the electrodes. f) Nyquist plots of a Zn‖Zn symmetric battery before cycling, and g) Electrolyte with the additive at various temperatures. h) Arrhenius curves of Zn‖Zn batteries at different temperatures.

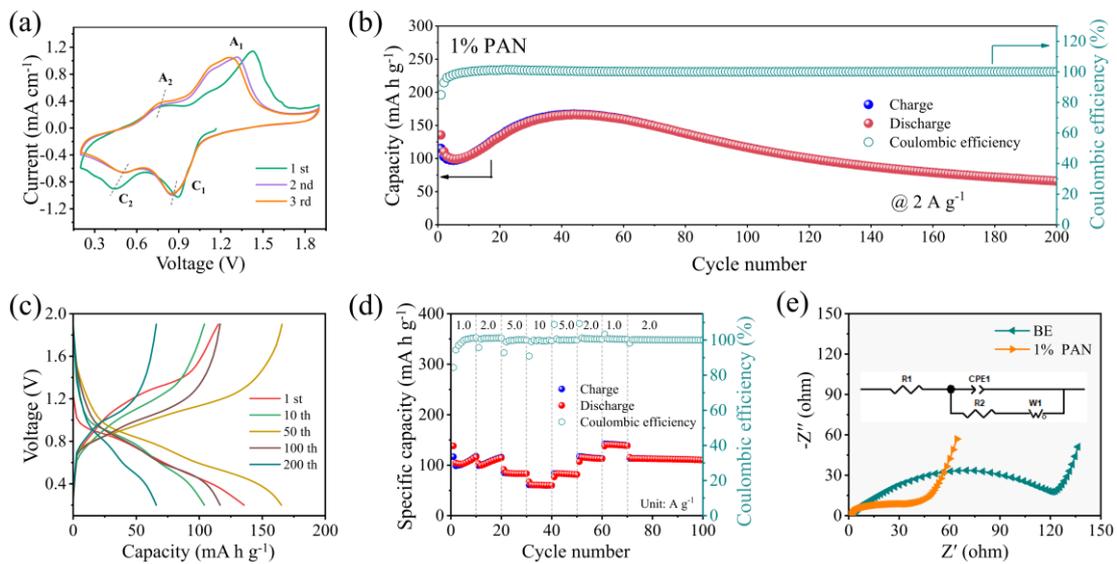

**Fig. 4**. a) Cyclic voltammetry curves of the first three cycles with a full battery of Zn‖ $V_2O_5$. b) Long cyclic performance at 2 A $g^{-1}$. c) Charge and discharge profiles at different cycles. d) Rate capabilities at the current densities of 1, 2, 5, and 10 A $g^{-1}$. e) EIS results of a Zn‖ $V_2O_5$ full battery before cycling.

**Supporting information for**

**Zincophilic armor: Phytate ammonium as a multifunctional additive for enhanced performance in aqueous zinc-ion batteries**

Fangyuan Xiao[a,b], Xiaoke Wang[a,b], Kaitong Sun[a], Qian Zhao[a], Cuiping Han[b,*], and Hai-Feng Li[a,*]

[a]*Institute of Applied Physics and Materials Engineering, University of Macau, Avenida da Universidade, Taipa, Macao SAR 999078, China*
[b]*Faculty of Materials Science and Energy Engineering/Institute of Technology for Carbon Neutrality, Shenzhen Institute of Advanced Technology, Chinese Academy of Sciences, Shenzhen Guangdong 518055, China.*

*Email addresses*: cp.han@siat.ac.cn (C. Han)
　　　　　　　　haifengli@um.edu.mo (H.-F. Li)

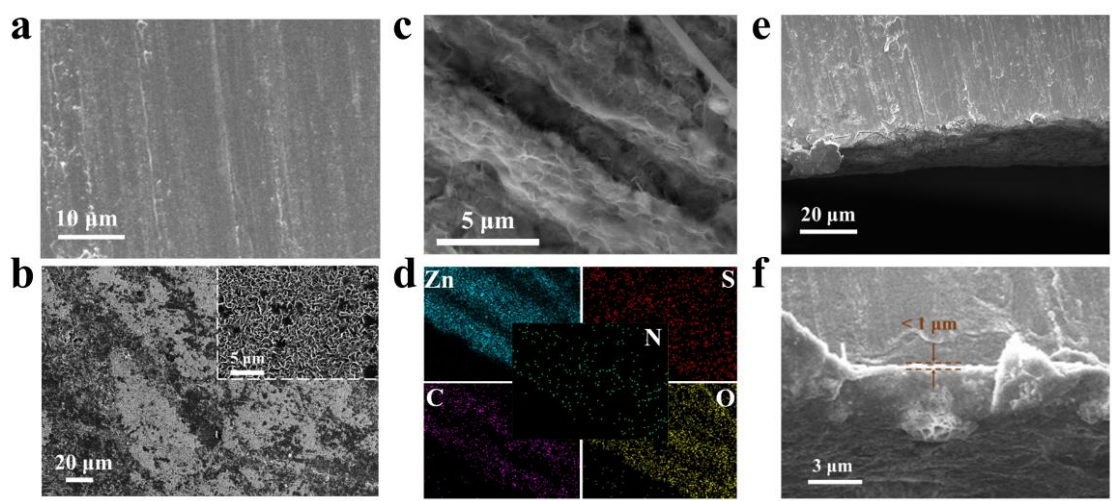

Figure. S1. SEM images of a) Bare zinc metal foil, b) Electrode paired with bare 1 M ZnSO$_4$ electrolyte after aging for 20 hours. c) Electrode with additive, and d) corresponding EDS mapping results. Cross-sectional images of the electrode: e) with bare electrolyte and f) with additive after 10 cycles at 1mA cm$^{-2}$, 1mA h cm$^{-2}$.

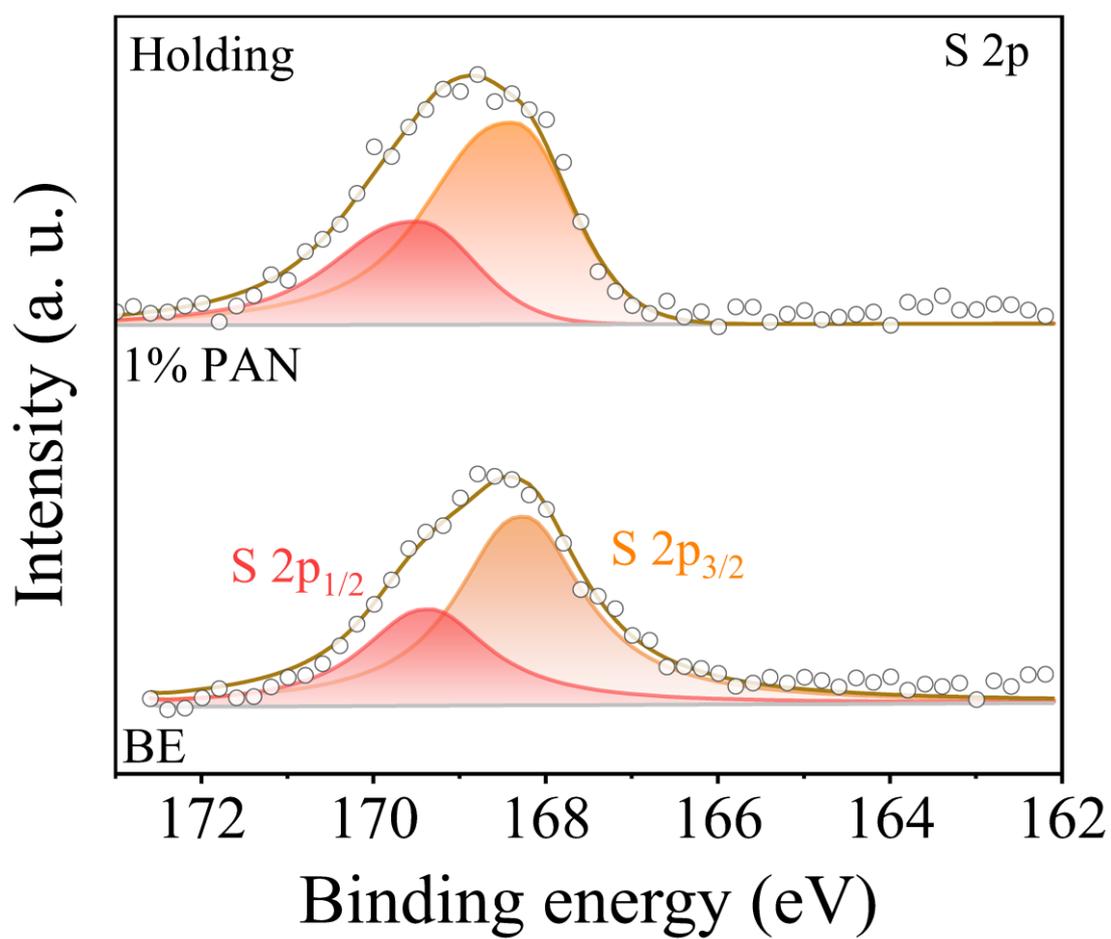

Figure. S2. Fine spectra around S 2p edge with XPS survey.

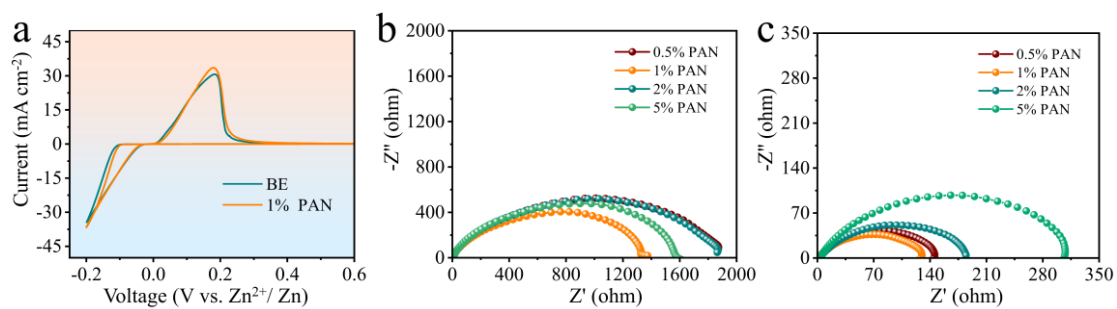

Figure. S3. a) Cyclic voltammetry (CV) curves of the cells before and after the addition of the additive with Zn‖ Ti asymmetry battery at a scan rate of 1 mV s⁻¹. b) Electrochemical impedance spectra results with different amounts of Zn‖ Zn symmetric battery assembled initially, and c) after 24 hours of aging.

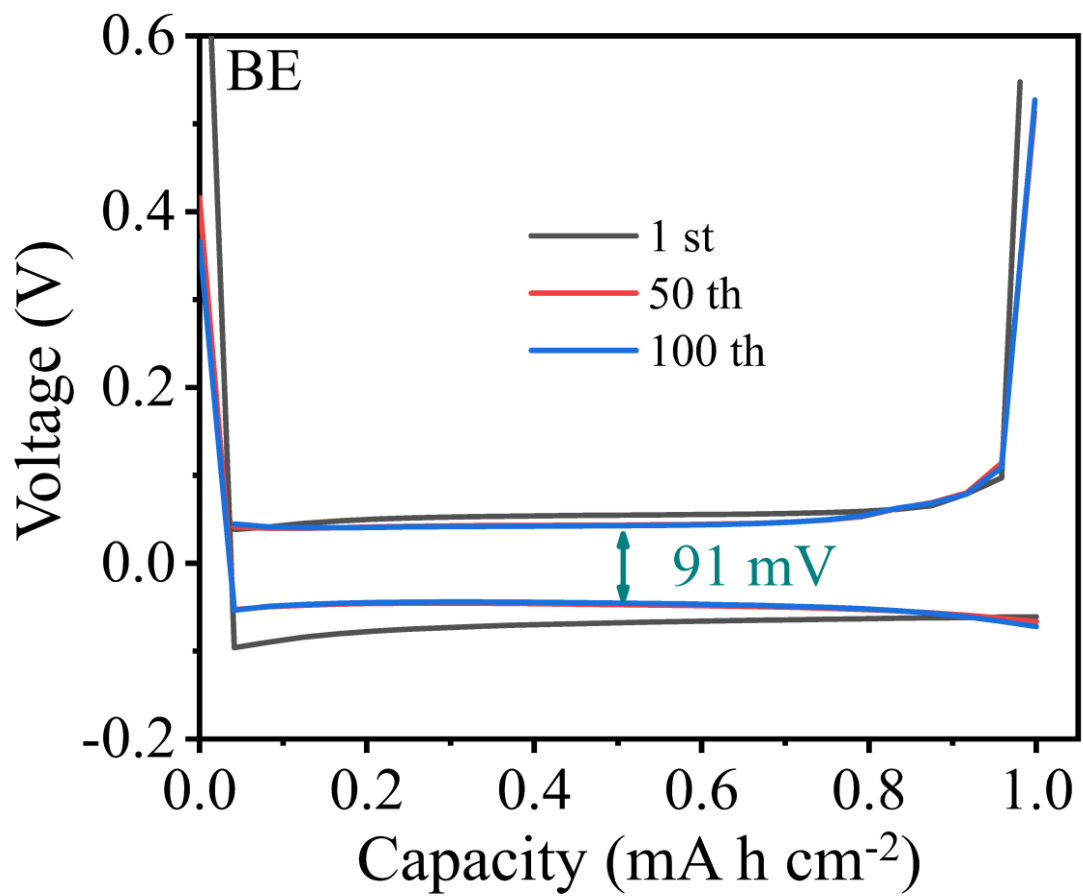

Figure. S4. Voltage- capacity curves on different cycles of bare electrolyte.

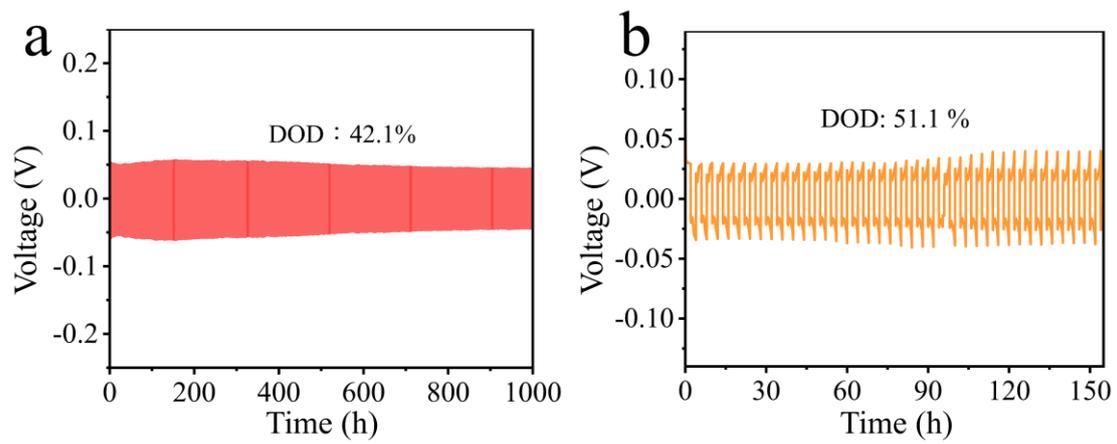

Figure. S5. Depth of discharge (DOD) test. a) DOD ~ 42.6%. b) DOD~ 51.1%

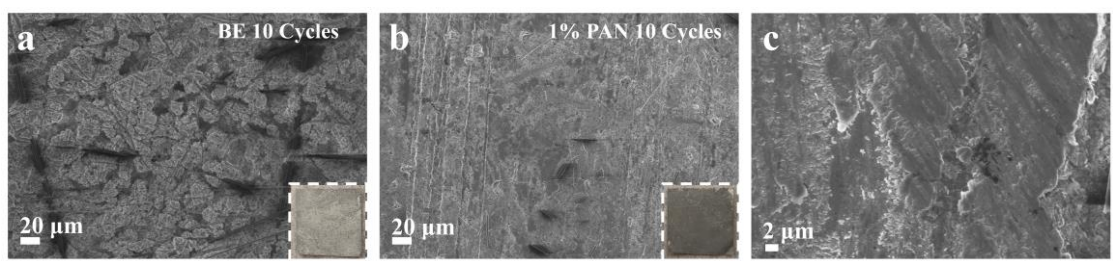

Figure. S6. Optical image (insertion) and SEM images of Zinc metal anode after 10 cycles at 1 mA cm$^{-2}$, 1mA h cm$^{-2}$. a) Bare electrolyte. b) With 1% volume PAN additive. c) High-magnification images of the sample with additive.

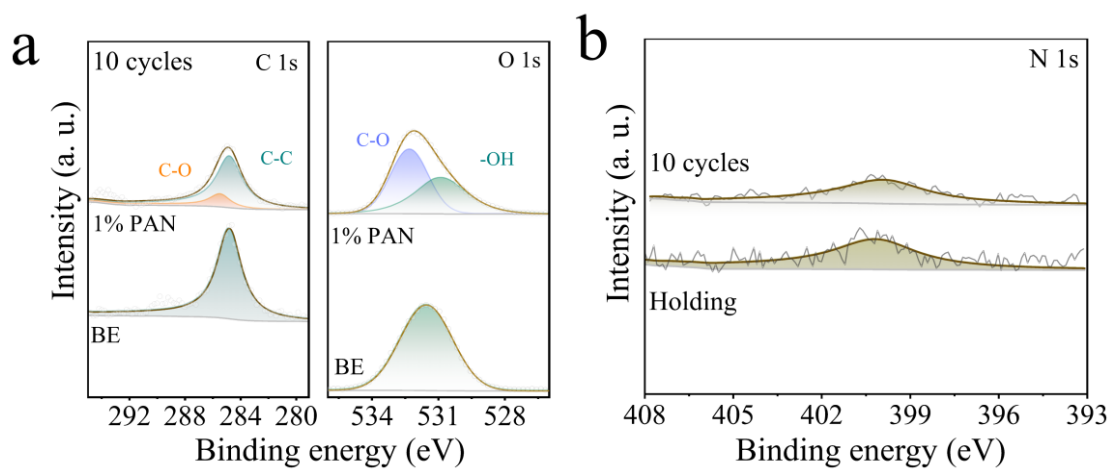

Figure. S7. XPS survey of the electrode. a) C 1s and O 1s edges before and after addition after 10 cycles at 1 mA cm$^{-2}$, 1mA h cm$^{-2}$. b) N 1s spectra after aging with 10 cycles.

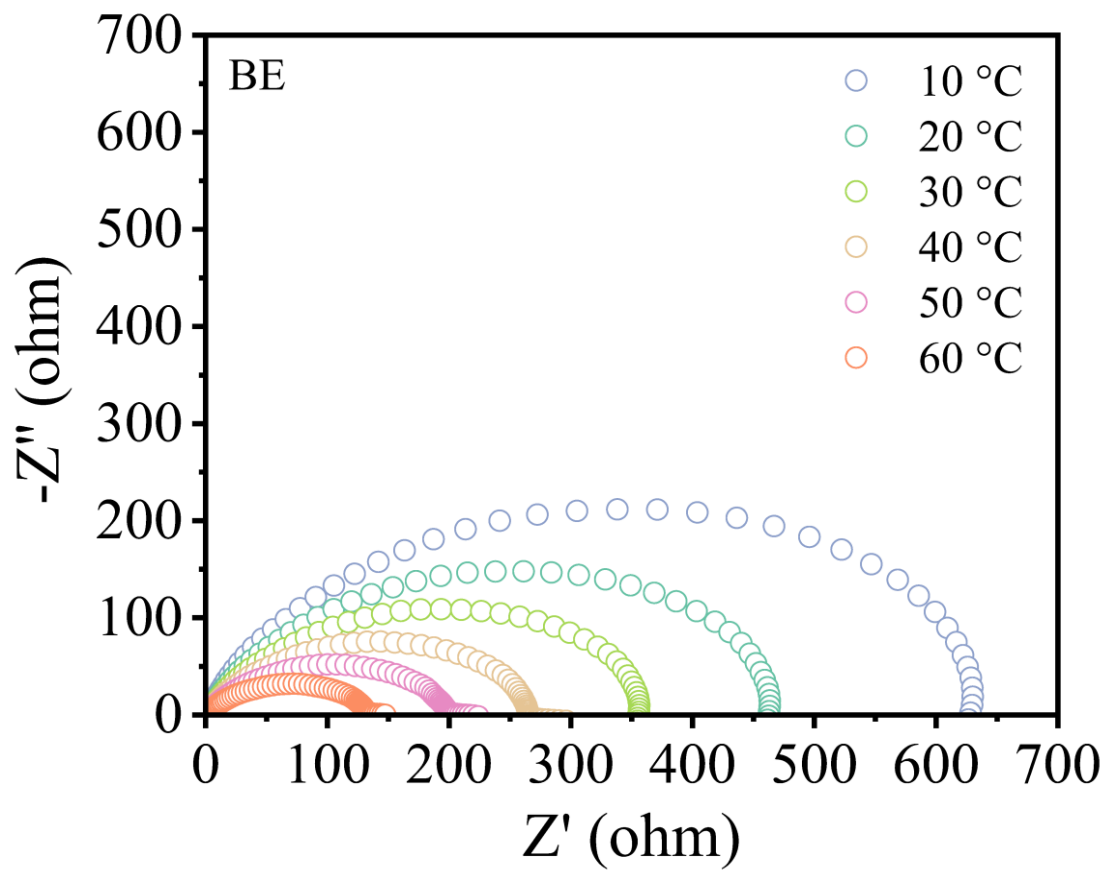

Figure. S8. Electrochemical impedance spectra (EIS) of bare electrolyte from 10 to 60 ℃ before cycling.